# Application of Machine Learning Method to Model-Based Library Approach to Critical Dimension Measurement by CD-SEM

P. Guo[1], H. Miao[1], Y.B. Zou[2#], S.F. Mao[3] and Z.J. Ding[4]*

[1]*CAS Key Laboratory of Mechanical Behavior and Design of Materials, Department of Modern Mechanics, University of Science and Technology of China, Hefei, Anhui 230027, China*

[2]*School of Physics and Electronic Engineering, Xinjiang Normal University, Urumqi, Xinjiang 830054, China*

[3]*Department of Engineering and Applied Physics, University of Science and Technology of China, Hefei, Anhui 230026. China*

[4]*Department of Physics and Hefei National Laboratory for Physical Sciences at Microscale, University of Science and Technology of China, Hefei, Anhui 230026, China*

#e-mail: zyb0617@mail.ustc.edu.cn

*e-mail: zjding@ustc.edu.cn

**Abstract**

The model-based library (MBL) method has already been established for the accurate measurement of critical dimension (CD) of semiconductor linewidth from a critical dimension scanning electron microscope (CD-SEM) image. In this work the MBL method has been further investigated by combing the CD-SEM image simulation with a neural network algorithm. The secondary electron linescan profiles were calculated at first by a Monte Carlo simulation method, enabling to obtain the dependence of linescan profiles on the selected values of various geometrical parameters (e.g., top CD, sidewall angle and height) for Si and Au trapezoidal line structures. The machine learning methods have then been applied to predicate the linescan profiles from a randomly selected training set of the calculated profiles. The predicted results agree very well with the calculated profiles with the standard deviation of 0.1% and 6% for the relative error distributions of Si and Au line structures, respectively. This result shows that the machine learning methods can be practically applied to the MBL method for the purpose of reducing the library size, accelerating the construction of the MBL database and enriching the content of an available MBL database.

Keywords：CD-SEM; MBL; Monte Carlo; machine learning; neural network

# 1 Introduction

The miniaturization of semiconductor devices has been the main direction of the device performance development in the past decades, and the number of transistors that can be accommodated on integrated circuits has been successfully doubled every two years in accordance with the Moore's law [1]. Even though Moore's law is now gradually failing in the post-Moore era at the atomic scale [2], the development of the semiconductor industry continues unabated. To achieve higher performance chips, the linewidth of semiconductors needs to be continuously reduced. Then the control of nanostructure dimensions is becoming more refined, and the critical dimension (CD) measurement technique becomes more important [3]. The CD measurement accuracy needs to be continuously improved, while the level of the CD measurement characterizes the level of development of the semiconductor industry.

Transmission electron microscopy [4], atomic force microscopy [5][6] and scanning electron microscopy (SEM) [7][8][9] are useful techniques for the CD measurement. Among them, only critical dimension scanning electron microscope (CD-SEM) is commonly adopted in industry for its fast and convenient measurement procedure. The use of secondary electrons as imaging signals has advantages on the CD measurement for both its sensitive response to the sample morphology and the high image resolution [10]. However, due to the edge effect of secondary electron emission [11][12][13] the linescan profile of the secondary electron signals has a certain spreading at an edge of the sample line structure [14][15][16]; hence, the accurate edge position is hardly to be determined from the secondary electron linescan profile directly. To cope with the edge effect problem in the CD measurement, some algorithms have been proposed especially for the line structures of large sizes above 100 nm. They are, such as, the maximum derivative method and curve fitting algorithm [17]. Novikov et al. [18] proposed a method to establish a simple correspondence between the sample model and the particular points (at the maximum and minimum values) of the secondary electron linescan curve. Frase et al. [19][20] proposed an exponential distribution fitting algorithm by fitting the linscan curve with some segmented continuous function by least squares. The threshold method [21] has been the main CD determination algorithm used in practice, with which the interval between the two points where the intensity corresponds to a fractional value of the maximum intensity of the secondary electron signals is determined as the CD value. However, these algorithms are purely empirical without a solid physical basis. Although they are simple to use, but the measurement accuracy is limited.

To achieve higher accuracy CD measurement, a model-based library (MBL) method has been proposed [22][23][24][25]. The MBL method determines the CD values of a nanostructure based on the physical principles of CD-SEM imaging. Using a Monte Carlo simulation method the electron scattering processes of an incident electron beam in the sample and the resultant generation, transmission and emission of the cascade secondary electron signals were simulated, enabling to calculate the linescan profile of the secondary electron signal intensity about the surface morphology [16][26][27]. By combing the Monte Carlo simulation of secondary electron signals with the sample geometrical structure modeling, a MBL database of CD-SEM linescan profiles for varied values of electron beam parameters and sample structural parameters can be constructed. Then a measured secondary electron linescan is matched with the closest one in the MBL database to deduce the structural parameter values of the sample. Li et al. [28] have characterized a Si line structure with the trapezoidal shape of a line cross section and investigated the effects of different parameters, such as, the linewidth and height, to the secondary electron linescan curves. Zou et al. [16] have further studied the MBL method to include those parameters considered later in ISO 21466; for example, the beam-sample interaction model contains two effective parameters, i.e. the line material and substrate material. Khan et al. [29] have applied the MBL method to characterize more complex sample structure having a smooth waveform and with a coated film, and we [30] have further extended the application of the method to rather more complex grating line in 3D morphology and in material components or electronic structure. Villarrubia et al. [31] have confirmed that the MBL method can be applied to linewidth measurement down to 10 nm scale; the measurements were in a good agreement with the direct observation by transmission electron microscope. Khan et al. [32] have investigated the influence the Monte Carlo model parameters to the CD determination and related theoretical uncertainty in the MBL method.

However, despite its clear physical background and higher measurement accuracy for smaller CDs the MBL method still suffers from a serious disadvantage of being computationally intensive as compared with the simple empirical methods. The first problem comes with the need of many parameters. According to ISO21466 [33], there are 8 and 11 independent geometrical parameters for a single and a double trapezoid model structures, respectively; and there are additional 3 and 4 parameters for specification of beam condition and probe, respectively. Usually for each of these parameters about 5-10 discrete values are necessary for construction of a useful database. Secondly, for each parameter value it is necessary to carry out a Monte Carlo calculation of a secondary electron linescan curve for a coordinate grid made of $\sim 10^2$ scanning points, and for each scanning point Monte Carlo simulation of electron trajectories is performed for at least $10^4$ incident primary electron trajectories and several tens of times the trajectories of the generated cascade secondary electrons. Therefore, a MBL database construction would cost a large amount of computation time as well as storage resources. At the same time, the search of the curves within such a big database in the matching process is also quite time consuming. One then needs to find the more economic approach towards the practical application of the MBL method.

The rapid development of machine learning methods in recent years has allowed the widespread applications in many fields, and deep learning is transforming the most fields of science and technology including also electron microscopy [34]. A typical application of deep learning is for reduction of the noise of an image [35][36]. Also semantic segmentation algorithms are often used in electron microscopy for automatic identification of local features [37][38] and automatic segmentation of electron microscopic images [39]. Tang and Spikes [40] used a neural network algorithm to quickly classify six page-rock samples based on SEM images. Dey et al. [41]

implemented unsupervised machine learning based on the extraction of SEM images from CD contour geometry, which can be quickly applied to the MBL database method with the existing mature neural network platform.

In this work, by combining with deep learning algorithms and using a neural network algorithm [42] we have enriched the MBL approach for the measurement of linewidth of the trapezoidal line shape. Guided by the ISO 21466, we at first built a 3D meshing for geometric modeling of a trapezoidal line structure. Then for different values of modeling geometric parameters, including lengths and angles, we have performed Monte Carlo simulations of secondary electron linescan curves to establish a MBL database. Taking a small number of data from the database to form a training set for machine learning, we have then predicted the linescan curves for other parameter values. The predicated linescan data are compared with the originally calculated data to test the predication accuracy, and an error distribution is then obtained. It is then found that the prediction error is quite small, indicating that a MBL can be firstly constructed in a small size with parse discrete values of parameters by using a Monte Carlo simulation method and later be extended very quickly and efficiently to a big size with dense discrete values of parameters by using a machine learning algorithm. This will save a big amount of computation time and also the data storage space while keeping the precision. On the other hand, by using sophisticated image recognition algorithms it is able to infer the sample geometry dimensions efficiently from the measured electron microscopic images and a MBL database. The present study thus expands the application scenarios of machine learning algorithms in scanning electron microscopy as well as to accelerate the application of MBL method in practice.

**2 Methods**

2.1 Monte Carlo model

The physical modelling of secondary electron signals is the theoretical basis of the MBL method. Monte Carlo simulation method is a mature and reliable technical tool for study complex electron interaction processes in a solid, and it is especially suitable for describing the mechanism of secondary electron generation, transport and emission processes [43][44][45]. The Monte Carlo method thus enables the image simulation of backscattered electron signals and secondary electron signals once a sample geometry is constructed [46][47][48][49].

For the elastic scattering of electrons in an atomic potential field the Mott's cross section [50] derived from relativistic quantum mechanics is used, which is more accurate than the classical Rutherford's cross section, particularly at low electron energies [51]. The Mott's differential scattering cross section is expressed as,

$$\frac{d\sigma_e}{d\Omega} = |f(\vartheta)|^2 + |g(\vartheta)|^2, \qquad (1)$$

where $\vartheta$ is the scattering angle; the direct scattering amplitude $f(\vartheta)$ and the spin reversal scattering amplitude $g(\vartheta)$ are obtained by solving the Dirac's equation with a partial wave expansion method. The scattering potential used in the present calculation follows the previous work [52][53][54] and contains two parts: the electrostatic interaction and the Furness-McCarthy exchange potential [55]. The Fermi distribution and the Dirac-Fock electron density [56] were used to describe the charge distribution of the nucleus and the electron cloud, respectively.

During inelastic scattering of an electron, both energy and direction of electron motion change, and the scattered electron loses kinetic energy to excite the electron degrees of freedom within the solid. The Bethe stopping power theory [57] describes the excitation of atoms and the averaged energy loss per unit distance of an electron; however, the theory does not address the detailed electronic excitation processes in solids and is not applicable in the low energy region. In order to deal reasonably with the complex and discrete electronic excitation processes in a solid, such as, single electron excitation and plasmon excitation, interband transition and inner-shell ionization within a dielectric function formalism, Penn has proposed to extrapolate an optical energy loss function, which is available from experimentally measured data, to the general energy loss function at finite momentum transfer [58]. Mao et al. used the full Penn algorithm to simulate the electron inelastic scattering process in a metal and the associated secondary electron emission process [59]. In the dielectric function formalism, the differential inelastic scattering cross section of electrons in a solid is written as,

$$\frac{d^2\lambda_{in}^{-1}}{d(\hbar\omega)dq} = \frac{1}{\pi a_0 E} \text{Im}\left\{\frac{-1}{\varepsilon(q,\omega)}\right\}\frac{1}{q}, \tag{2}$$

where $a_0$ is the Bohr radius, $\hbar$ is the reduced Planck constant; $\varepsilon(q,\omega)$ is the complex dielectric function of the medium, and $\text{Im}\{-1/\varepsilon(q,\omega)\}$ is called the energy loss function which determines completely the probability of electron inelastic scattering with an energy loss $\hbar\omega$ and momentum transfer $\hbar q$. The inelastic mean free path, $\lambda_{in}$, is then obtained as,

$$\lambda_{in}^{-1} = \int_0^{E-E_F} d(\hbar\omega)\int_{q_-}^{q_+} dq \frac{d^2\lambda_{in}^{-1}}{d(\hbar\omega)dq}, \tag{3}$$

where $E_F$ is the Fermi energy; the upper and lower limits for the momentum integration are $q_\pm = \left(\sqrt{2E} \pm \sqrt{2(E-\hbar\omega)}\right)/\hbar$. According to Penn [58], the optical energy loss function, $\text{Im}\{-1/\varepsilon(q=0,\omega)\}$, can be extended to the $(q,\omega)$-space by

$$\text{Im}\left\{\frac{-1}{\varepsilon(q,\omega)}\right\} = \int_0^\infty d\omega_p g(\omega_p) \text{Im}\left\{\frac{-1}{\varepsilon_L(q,\omega;\omega_p)}\right\}, \tag{4}$$

where $\omega_p$ is the plasmon frequency; the expansion coefficient $g(\omega_p)$ is related to the optical energy loss function at $q=0$ and is given by,

$$g(\omega) = \frac{2}{\pi\omega}\text{Im}\left\{\frac{-1}{\varepsilon(q=0,\omega)}\right\}. \tag{5}$$

$\varepsilon_L(q,\omega;\omega_p)$ is the Lindhard dielectric function whose real and imaginary parts are, respectively, expressed as,

$$\mathrm{Re}\{\varepsilon_L(q,\omega;\omega_p)\} = 1 + \frac{2}{\pi a_0 q}\frac{1}{Z}\left\{\frac{1}{2} + \frac{1}{8Z}F\left(Z - \frac{X}{4Z}\right) + \frac{1}{8Z}F\left(Z + \frac{X}{4Z}\right)\right\}, \tag{6}$$

$$\mathrm{Im}\{\varepsilon_L(q,\omega;\omega_p)\} = \begin{cases} \dfrac{1}{8a_0 k_F}\dfrac{X}{Z^3}, & 0 \le X \le 4Z(1-Z) \\ \dfrac{1}{8a_0 k_F}\dfrac{1}{Z^3}\left\{1 - \left(Z - \dfrac{X}{4Z}\right)^2\right\}, & |4Z(1-Z)| \le X \le 4Z(1+Z) \\ 0, & \text{otherwise} \end{cases} \tag{7}$$

where $k_F$ is the Fermi wavevector, $X = \hbar\omega/E_F$, $Z = q/2k_F$ and

$$F(x) = (1 - x^2)\ln|(x+1)/(x-1)|. \tag{8}$$

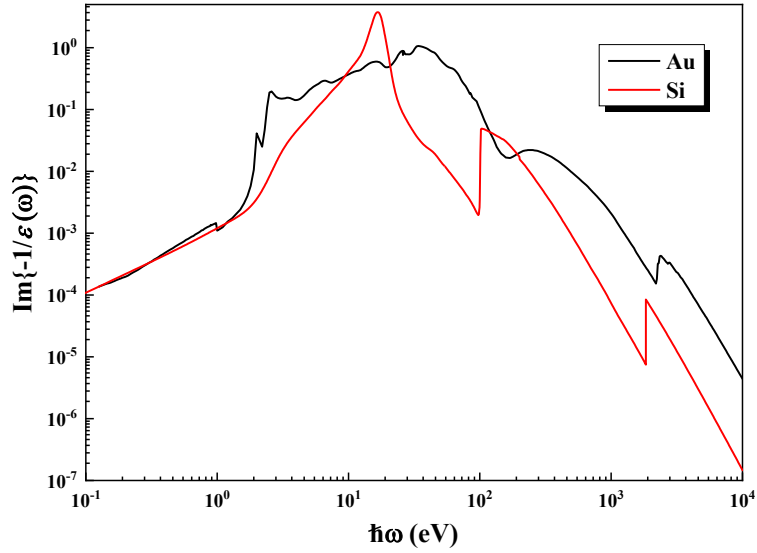

Fig. 1 Optical energy loss function of Au and Si.

The optical energy loss function is an important physical quantity for the calculation of inelastic scattering cross section. Its accuracy will directly affect the calculation results of secondary electron yields. Fig. 1 shows the optical energy loss function data for Au and Si used in this work. For Au the low-energy ($\hbar\omega < 94.52$ eV) data are taken from the Palik's database [60] and the high-energy data are derived from the atomic scattering factor [61]. For Si the low-energy ($\hbar\omega < 200$ eV) data are derived from the reflected electron energy loss spectrum [62] by using a reverse Monte Carlo analysis [63], and the high-energy data are calculated from the atomic scattering factor [61].

2.2 Machine Learning

Although the MBL method has high measurement accuracy, it requires large scale calculations to establish a database. In order to reduce the computation cost for simulating the secondary electron linescan profiles, this paper uses the neural network method to replace in part the direct Monte Carlo simulation process. By calculating a small amount of secondary electron linescan curves as a training set, then a set of reliable neural network models are trained to predict the profiles for other structural parameter values. In this way one can speed up the construction of the MBL database.

Machine learning has been extensively applied in various cross-disciplinary fields, especially the popularity of various neural network algorithms enables more and more fast prediction of physical results. The neural net adopted in this paper is a fully connected neural network based on the keras platform [63]. Fig. 2 illustrates the basic fully connected neural network model in three layers consisting of input layer, output layer and hidden layer. The input layer of this network structure has $d$ neurons, the hidden layer has $q$ neurons, and the output layer has $l$ neurons. The weight from the $i$th neuron in the input layer to the $h$th neuron in the hidden layer is $v_{ih}$, and the weight from the $h$th neuron in the hidden layer to the $j$th output layer is $\omega_{hj}$, the bias of the $j$th neuron in the output layer is $\theta_j$. A training set is denoted by $D = \{(\mathbf{x}_1, \mathbf{y}_1), (\mathbf{x}_2, \mathbf{y}_2), ..., (\mathbf{x}_m, \mathbf{y}_m)\}, (\mathbf{x}_i \in \mathbb{R}^d, \mathbf{y}_i \in \mathbb{R}^l)$, where the input vector $\mathbf{x}$ represents the structural parameters of a trapezoidal line, and the output vector $\mathbf{y}$ is the secondary electron linescan profile and the components are the intensities at each scanning point.

There are two main points in the fully connected neural networks. First, the sample input is operated from left to right to derive the output of the neural network, and this process is called forward propagation. Second, a loss function is defined for the measurement of the deviation between the output value of the network and the true value of the sample. Then, through the idea of gradient descent, from the right to the left, the loss function is allowed to find the bias derivative for each weight and bias, where the bias derivative is used to update the weights and bias step by step, and this process is called backward propagation.

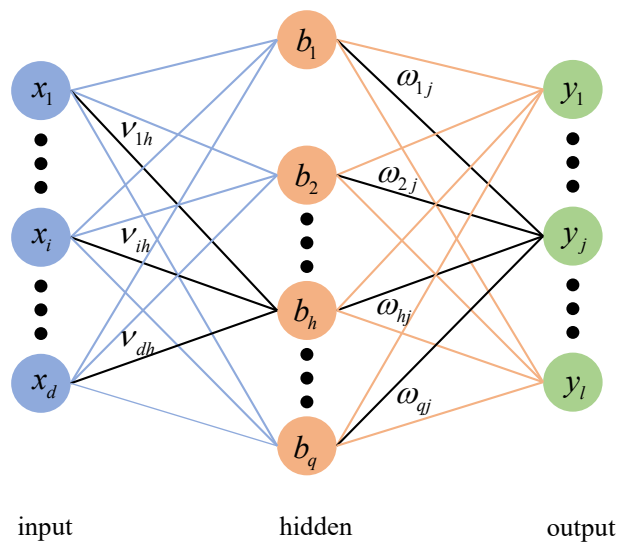

Fig. 2 Schematic diagram of fully connected neural network structure.

The neuron activation functions of both the hidden layer and output layer use the Sigmoid function. For a training sample $(\mathbf{x}_k, \mathbf{y}_k)$ it is easy to derive its output as

$$\hat{y}_j = f(\beta_j - \theta_j), \tag{9}$$

where $\beta_j = \sum_{h=1}^{q} \omega_{hj} b_h$ is the input of the $j$th output neuron. Then a loss function is defined as

$$E_k = \frac{1}{2} \sum_{j=1}^{l} (\hat{y}_j^k - y_j^k)^2, \tag{10}$$

where $y_j^k$ is the real sample output result, $\hat{y}_j^k$ is the prediction.

The error back propagation algorithm is based on a gradient descent strategy to adjust the parameters in the direction of the negative gradient of this loss function $E_k$,

$$v \leftarrow v + \Delta v; \tag{11}$$

$$\omega \leftarrow \omega + \Delta\omega. \tag{12}$$

For an error $E_k$, given the learning rate $\eta$ (the step size of each iteration), it follows that,

$$\Delta\omega = -\eta \frac{\partial E_k}{\partial \omega_{hj}}. \tag{13}$$

According to the chain rule,

$$\frac{\partial E_k}{\partial \omega_{hj}} = \frac{\partial E_k}{\partial \hat{y}_j^k} \cdot \frac{\partial \hat{y}_j^k}{\partial \beta} \cdot \frac{\partial \beta}{\partial \omega_{hj}}, \tag{14}$$

and by the definition of $\beta_j$, the last term in Eq. (14) can be deduced $\frac{\partial \beta_j}{\partial \omega_{hj}} = b_h$. Simoid function has the following property,

$$f'(x) = f(x)(1 - f(x)). \tag{15}$$

Based on Eqs. (9) and (10), the product of the first two terms in Eq. (14) can be deduced as

$$g_j = \frac{\partial E_k}{\partial \hat{y}_j^k} \cdot \frac{\partial \hat{y}_j^k}{\partial \beta} = \hat{y}_j^k \left(1 - \hat{y}_j^k\right)\left(y_j^k - \hat{y}_j^k\right). \tag{16}$$

The updated values of the key parameters of the algorithm can be derived as follows,

$$\Delta\omega_{hj} = \eta g_j b_h. \tag{17}$$

Similarly, the parameters $\Delta\theta_j$ and $\Delta v_{ih}$ can be updated.

In this work, the input vector $\mathbf{x}$ is composed of the structural parameters (i.e. top CD, height and sidewall angle for a single layer trapezoid in Fig. 3, and in addition the chamfer radii and an additional sidewall angle for a double layer trapezoid in Fig. 7) and the dimension is small. But the dimension of the output vector $\mathbf{y}$ is large; here we adopt a uniform scanning grid, thus the information of the horizontal coordinates can be omitted and we keep only the intensity values of the linescan at the scanning grid points. With the above mentioned error back propagation algorithm, a fully connected neural network model can be trained to meet the requirements for constructing the correspondence between the structure and the linescan curve.

## 3. Results and Discussion

3.1 Single layer trapezoidal line

The required input parameters in the Monte Carlo simulation for building a MBL database include: 1) sample geometric parameters (top CD, height, sidewall angle, roundness of the top and bottom corners of a trapezoidal line structure) whose values are varied; 2) sample material parameters (density, optical constants and work function) whose values are usually certain or reasonably chosen for a given sample configuration; 3) electron beam parameters (primary energy, incidence angle and beam spot size) whose values are varied; 4) signal detection parameter (a binary value) which accounts for the electric field applied to the detector. Some other parameters in the Monte Carlo simulation, like the linescan range and step size, are not the real variables of a MBL database. Those helpful parameters that have significant impacts on the linescan result are considered in the present calculations to ensure that the MBL database is enough comprehensive and includes significantly varied secondary electron linescan profiles.

Fig. 3 shows the structure of a trapezoidal line made of Au element. The geometric structural parameters include the top CD ($T$), height ($H$) and sidewall angle ($\theta$) of the trapezoid. We set $T$ varied from 10 nm to 100 nm with an interval of 5 nm for a total of 19 values; $H$ varied from 10 nm to 50 nm with an interval of 5 nm, for a total of 9 values; and $\theta$ varied between 0 and 30° with an interval of 5°, for a total of 7 values. A secondary electron linescan curve $\{I_i, i=1,2...N\}$ is a $N$-dimensional vector, where $N$=200, with the position pixel range of 200 nm at an interval of 1 nm. A Gaussian electron beam of 1 keV with a beam diameter of 5 nm is incident vertically onto the sample surface; the number of incident electrons is 20,000 at each scanning point. The simulation is performed with our up-to-date Monte Carlo simulation code, CTMC-3DSEM [65].

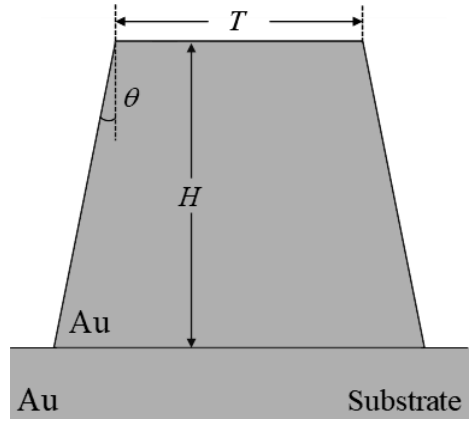

Fig. 3 Geometry of a single layer trapezoidal line structure.

The Monte Carlo simulated secondary electron linescan curves, $I^{MC}$, for different values of geometrical parameters of the gold line structure are shown in Fig. 4. It can be seen that the linescan profile intensity blooms at an edge of the trapezoidal line; this is just the so-called edge effect of the secondary electron emission. The position of the maximum intensity representing approximately the edge position is certainly related to the $T$ value. When $T$ is increased within a narrow range from 20 nm to 80 nm in Fig. 4(a), the peak position moves linearly, while the shape and the peak intensity of the linescan remain almost unchanged. Then the change of the peak position accurately reflects the change of the trapezoidal linewidth. However, because of the finite peak width, it is hard to accurately decide the linewidth directly from the peak position. In the MBL method, the entire curve shape within a certain range of the intensity bloom is matched with a SEM observation, rather than simply relying on the location of a single particular point in the profile; in this way the measurement accuracy can be guaranteed. As the height of the trapezoidal line increases in Fig. 4(b), the spreading of the peak region increases slightly; while as the sidewall angle increases in Fig. 4(c), the linescan curve at the waist of the trapezoidal slopes changes slowly and the spreading of the peak region increases significantly. Although the trends of curve changes in Fig. 4(a) and Fig. 4(c) are similar, there are differences in the line shapes in the two cases, and the best matching of the MBL database could enable to distinguish them.

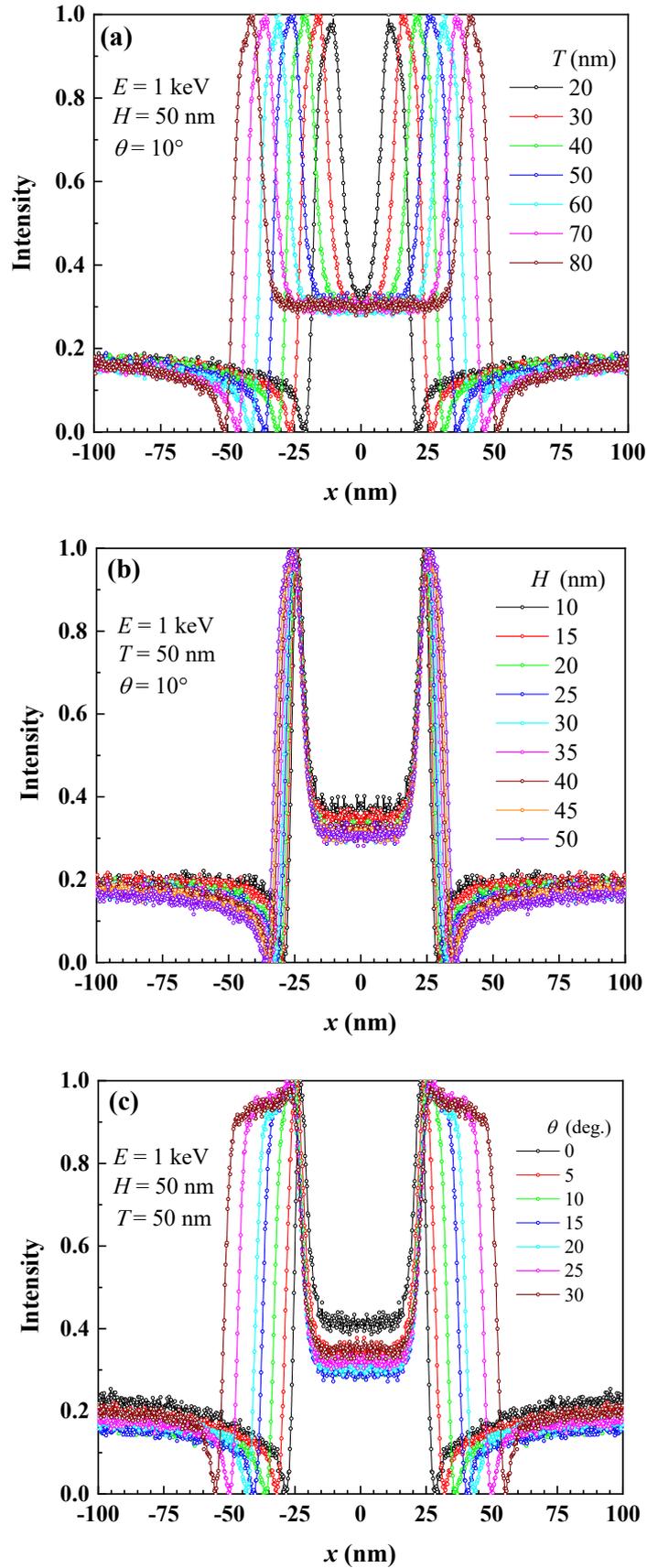

Fig. 4 Normalized linescan curves simulated by a Monte Carlo method for an Au line structure and for different values of structural parameters: (a) top CD; (b) height; (c) sidewall angle.

In order to accelerate the construction of a MBL database by saving the workload on the computation of linescan curves, the keras-based fully connected neural network method is used in this work to train the model for the secondary electron linescan curves and the corresponding geometric structure parameters. The training set is taken from the existing linescan calculation results $I^{MC}$, in which there are 1197 data for the gold trapezoidal line structure; 65% of which are divided into the training set and the remaining 35% into the test set. The input parameters are three independent parameters, i.e. $T$, $H$ and $\theta$, of a trapezoid for description of the geometric structure dimensions of the trapezoidal line structure. A fully connected neural network is used to build a 7-layer network structure, where the first layer is the input layer containing 3 input neurons, and the second to the sixth layers are the five hidden layers. The number of neurons from the lower layer to the higher layer of the network gradually increases, i.e. it is 8, 16, 32, 64 and 256 from the second to the sixth layers. The seventh layer is the output layer containing 200 output neurons. The learning rate is set to 0.0002, and the training batch is a group of 10 samples for 500 rounds.

The model trained according to the training neural network can predict the calculation results very quickly. Fig. 5 shows a comparison of the directly calculated Monte Carlo results $I^{MC}$ with the predication results $I^{ML}$ by the trained machine learning model for four cases of randomly selected parameter sets. It can be seen that the accuracy of the prediction results is very high. In addition, the computation speed is very fast. The direct Monte Carlo simulation for constructing a MBL database, by tracking each electron interaction with the solid, sampling secondary electron generation, their transport trajectories and escape statistics, consumes heavily computing resources. For example, the calculation costs ~100 hours with 10 nodes or 400 cores in a supercomputer. If one needs to increase the number of incident electrons to reduce the statistical fluctuation of the calculated linescan curve and/or to further refine the coordinate mesh, the computational cost would be correspondingly multiplied. However, for 500 times of model training by a neural network it takes only about 2 min in a personal laptop and the obtained model file is less than 20MB in size, which greatly reduces the storage cost. Especially, for expanding the database by further parameter partitioning in future the storage space will be expanded accordingly for the direct Monte Carlo simulated data. However, once the neural network structure is determined it will almost not increase the memory space for an already established MBL database while the effective parameter value sets is expanded. In addition, the neural network model prediction of a linescan curve consumes time less than 1 s, which greatly saves the necessary computation time by a direct Monte Carlo simulation while still maintaining the enough accuracy.

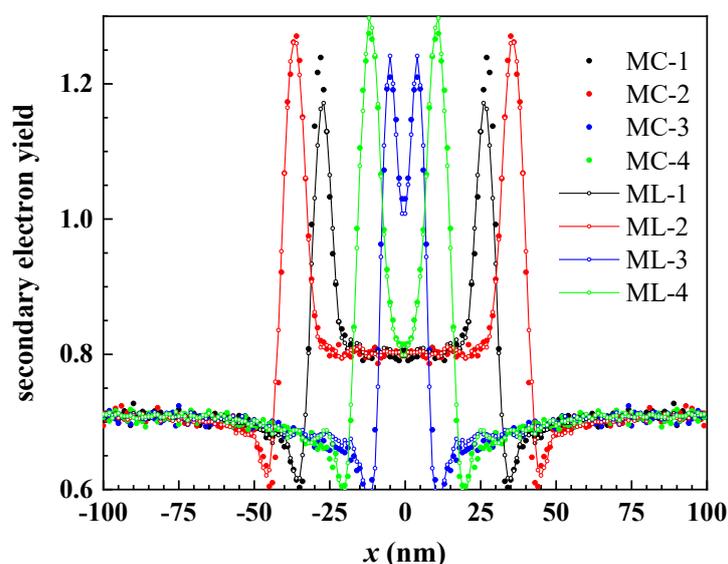

Fig. 5 Comparison of the predicted linescan curves $I^{ML}$ (empty circles and solid lines, ML) by the machine learning method with the direct Monte Carlo simulated data $I^{MC}$ (solid circles, MC).

The predicted linescan curve by the neural network model has also $N$ (=200) points, and the relative difference between the predicted data $I^{MC}$ and the directly simulated data $I^{MC}$ is defined as the error. The root mean square error (RMSE) of a linescan curve is then obtained as,

$$\mathrm{RMSE} = \sqrt{\frac{1}{N-1}\sum_{i=1}^{N}\left\{\left(I_i^{MC}-I_i^{ML}\right)\Big/I_i^{MC}\right\}^2}\ . \tag{18}$$

Fig. 6(a) shows the RMSE distribution for the 427 test samples we selected; the maximum relative error is less than 6%. The cumulative distribution function in Fig. 6(b) indicates that the predicted results within RMSE of 3% account for 94.1% of the total predicted results. This verifies that the trained model has enough accuracy.

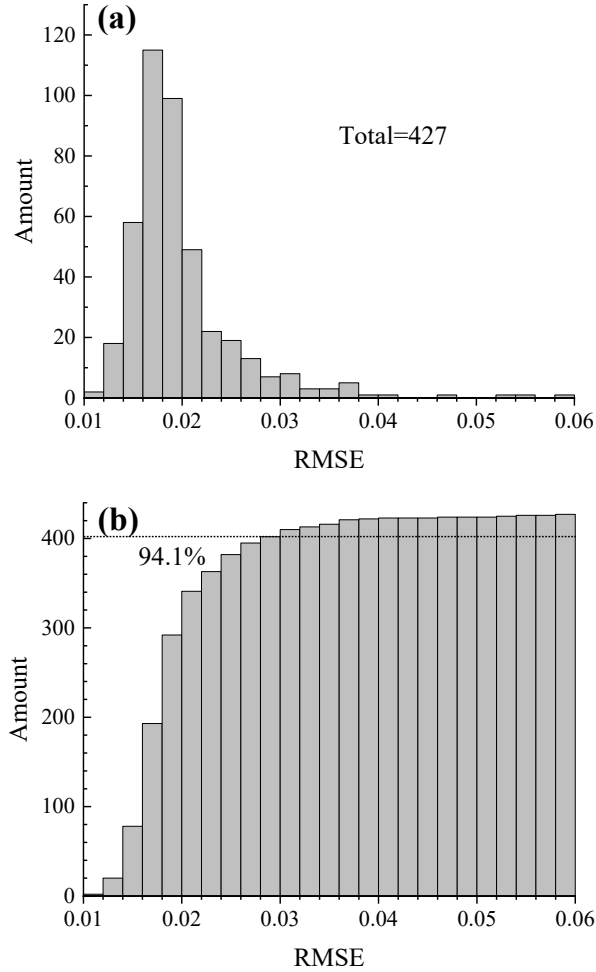

Fig. 6 (a) The RMSE distribution of the predicted results by the machine learning method; (b) the cumulative distribution function of RMSE.

3.2 Double-layer trapezoidal line

Using the same Monte Carlo model, we have also performed calculation for a Si trapezoidal line. To be better fit the actual application scenario, a symmetrical double-layer trapezoidal line structure is constructed. Taking into account the parameter setting according to ISO 21466 [33], the rounded edge corners are treated. In addition, to fully consider the influence of nearby line structures to the secondary electron emission via absorption of emitted electrons, three equivalent nanolines in a pitch of 90 nm are constructed and the linescan profile over the central line is calculated. Fig. 7. shows the modeled double-layered trapezoidal structure. The value of top CD, $T$, is taken as 30-75 nm at an interval of 5 nm, for a total of 10 values; the value of upper trapezoidal height, $H^t$, is taken as 0-40 nm at an interval of 10 nm, for a total of 5 values; the value of lower trapezoidal height, $H^b$, is taken as 30-80 nm at an interval of 10 nm, for a total of 6 values; the value of top sidewall angle, $\theta^t$, is taken as 0-8° at an interval of 2°, for a total of 5 values, and the same division is also applied to the bottom sidewall angle, $\theta^b$, for a total of 5 values; the radius of the top arcs and bottom arcs are taken the same, $R^t = R^b$, and the value of is 0, 5 and 10 nm, for a total of 3 values. The total number of samples constructed is thus 22,500. A Gaussian electron beam of 1 keV with a beam diameter of 5 nm is incident vertically onto the sample surface; the number of incident electrons is 20,000 at each scanning point, and the number of scanned pixel points is 300. The simulation is performed by using a parallel computer with 10 nodes and 40 cores per node. It takes about 8 min to compute a sample task using 400 cores in parallel, and the total task has costed ~20 days in a supercomputer.

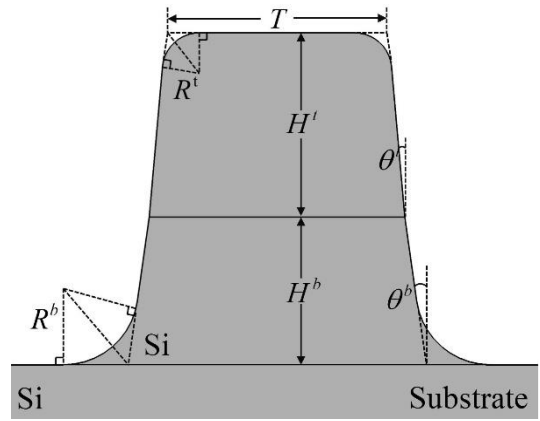

Fig. 7 Geometry of a double-layer trapezoidal line structure.`

Fig. 8 shows the secondary electron linescan curves of the Si double-layer trapezoidal line nanowire. It can be seen that there is an intensity bloom in the linescan curve at an edge of the line in similar to the case of the trapezoidal line in Fig. 4. In Fig. 8(a) the line structure is a rounded rectangle as the sidewall angles are taken as 0; with the increase of the corner arcs radius, the peak region of the linescan curve is less changed while the intensity around the valley region is reduced. In Fig. 8(b) with the increase of the bottom sidewall angle the peak region of linescan curve is gradually expanded but the peak position changes little. In Fig. 8(c) the peak position changes linearly with top CD, while the shape of the linescan signal curve and the peak intensity remain almost unchanged. With the increase of the top layer height of the trapezoidal line, the spreading of the peak area increases significantly as shown by Fig. 8(d).

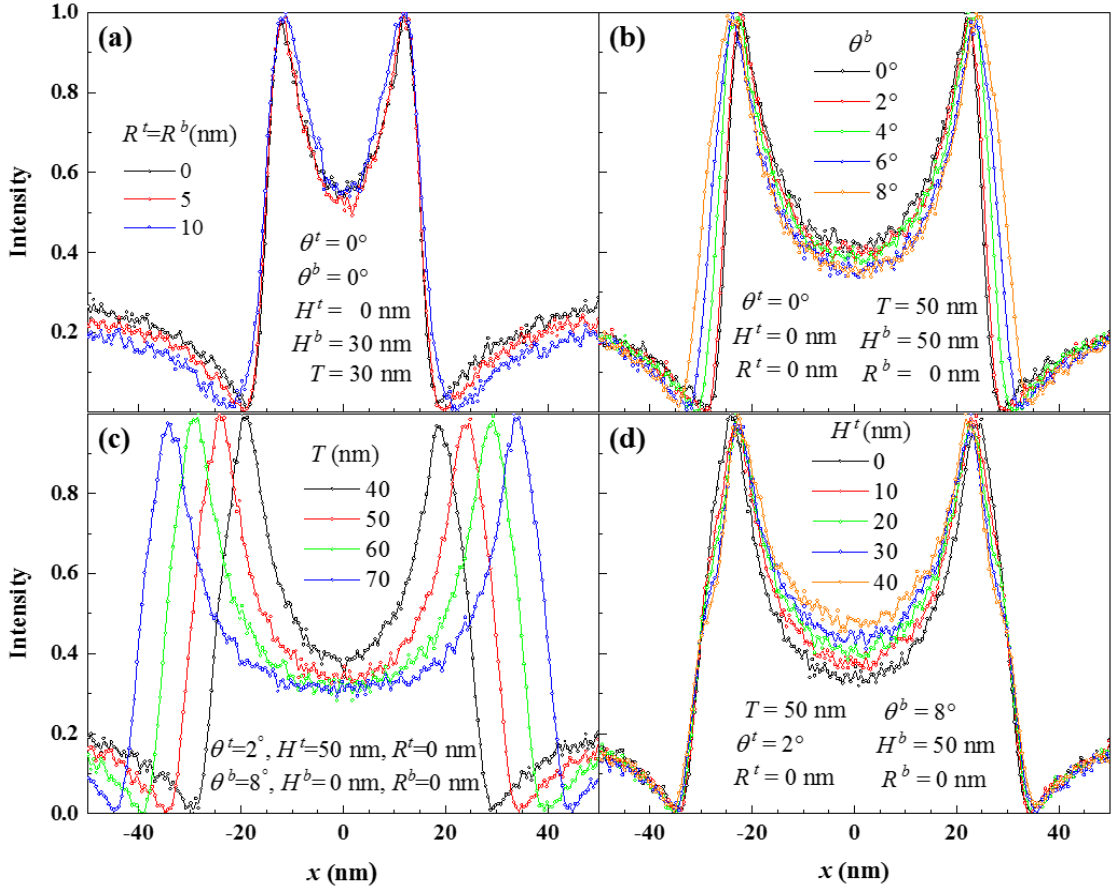

Fig. 8 Normalized linescan curves simulated by a Monte Carlo method for a Si double-layer trapezoidal line structure at different values of structural parameters: (a) radius of the arcs; (b) bottom sidewall angle; (c) top CD; (d) top trapezoidal height.

For this double-layered trapezoid line structure, more structural parameters are necessary and, hence, a direct Monte Carlo simulation of the linescan curve needs to consume much more computation time. The machine learning is expected to be more useful in this case for MBL database extension. We then use the neural network method to train the model, and to predict the linescan curves of other geometric parameters. The prediction results are also compared with the simple interpolation method to demonstrate the advantage of the machine learning method.

For the double-layer trapezoidal structure, although the the radius of the top arcs and bottom arcs are taken the same there are still a total of 22,500 linescan curves by a direct Monte Carlo calculation. Among them we randomly selected 5000 different parameter combinations for training, and then we can predict the remaining 17,500 cases to be compared with direct Monte Carlo calculations. In addition, we also compare with the simple interpolation method. A similar method is used to construct a fully connected neural network. The training set with 5000 data for this double-layer structure is much greater than the previous one, 778, for the single-layer structure. We still set up a 7-layer network structure. The difference from the previous one is that the input layer has 6 neurons, and the number of neurons in the hidden layer has increased. The number of neurons in the second to fifth layers is set to 16, 64, 128 and 256, the sixth layer is still set to 256 neurons. The final output layer keeps the same number of pixels, 300, in a linescan.

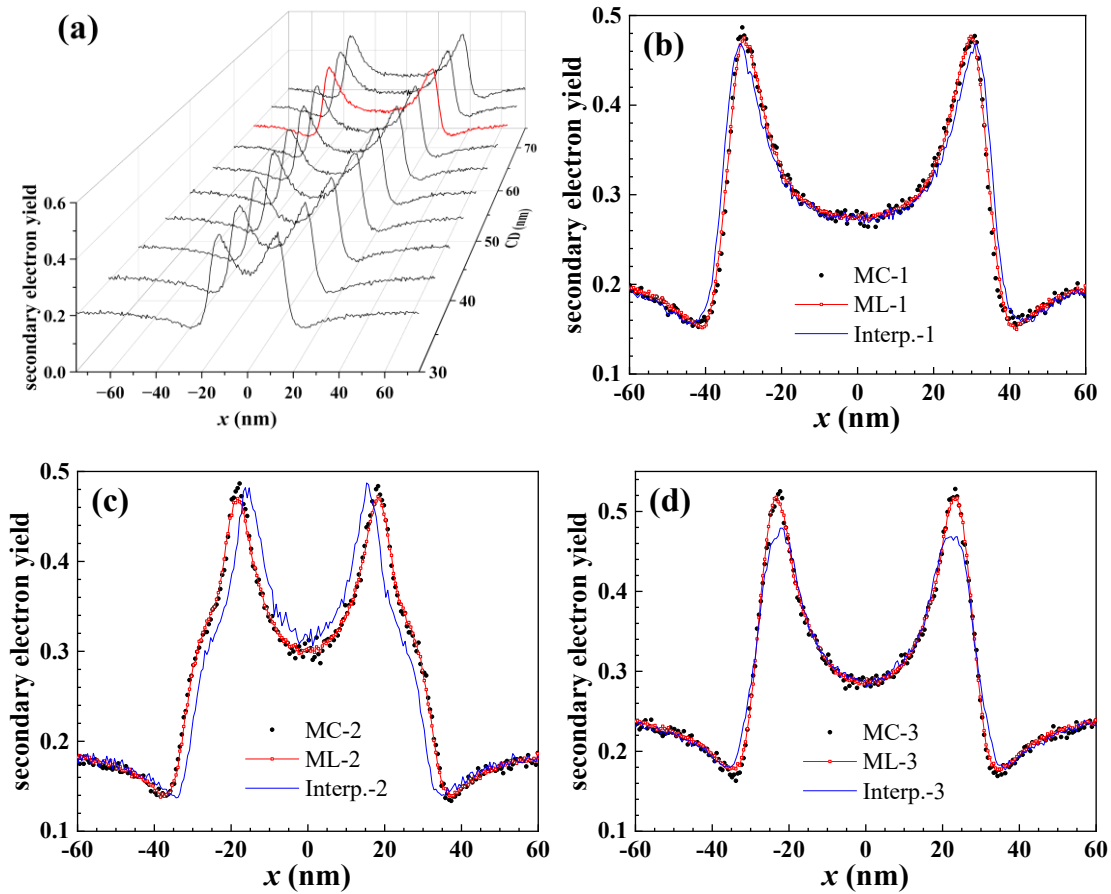

Fig. 9 (a) Example of the direct Monte Carlo calculated linescan curves (black lines) for a Si double-layer trapezoidal line structure. The red line is the one to be used in (b) for comparison. (b)-(d): Comparisons between the machine learning predicted linescan curve (empty circles and red line), the interpolation (blue line) and the direct Monte Carlo simulation (solid circles) for three different parameter sets.

Fig. 9(a) shows the simulation results for a group of $T$ varied from 30 to 75 nm at an interval of 5 nm, while other parameters are unchanged: $\theta^t = 4°$, $\theta^b = 8°$, $H^t = 20$ nm, $H^b = 30$ nm, and $R^t = R^b = 5$ nm. In analogous to Fig. 5, Fig. 9(b)-9(d) demonstrate 3 cases of comparison made between the direct Monte Carlo simulation, machine learning predicted result and simple interpolation. As it can be seen that in all the cases shown, the machine learning predication agrees excellently with the Monte Carlo simulation while interpolation has certain discrepancy: the peak position in Fig. 9(c) and the peak height in Fig. 9(d) change significantly. Therefore, the accuracy of neural network method is much better than that of the interpolation method.

In order to further explore the accuracy and generalization ability of the neural network model, we have calculated the RMSE of the machine learning prediction and the Monte Carlo simulation according to Eq. (18). Fig. 10(a) shows the RMSE distribution of the prediction results for $T$ in the range of 30-65nm. It can be seen from the cumulative function of RMSE that the RMSE within 0.07% is more than 94.85% of the cases, as shown in Fig. 10(b); the predicted result is very satisfactory. When we apply the model for predication of the data out of range, e.g. for $T$=70 nm in Figs. 10(c) and 10(d), the RMSE is generally below 10% and RMSE within 6.25% is more than 94.53% of the cases; the predication thus still maintains a high accuracy. Fig. 10(e) and (f) are for the case of $T$=75 nm, the prediction accuracy is decreased as the distance from the training set is further away. The RMSE of prediction within 11.5% ia more than 93.77% of cases.

Through the above comparison, it can be seen that the accuracy and the generalization ability of the prediction results of the machine learning method are excellent. Using only 22% of the data set, the remaining 78% of the test data can be quickly and accurately predicted. Compared with Monte Carlo simulation which took ~3 months for calculation of the 17,500 data, a trained neural network only takes ~1 minute. Then by the one-to-one correspondence between the sample structure and the linescan curve, the CD of a nanometer line structure can be obtained by matching a measured linescan curve with from the machine learning predicated in an extended MBL database of the Monte Carlo simulated linescan curves.

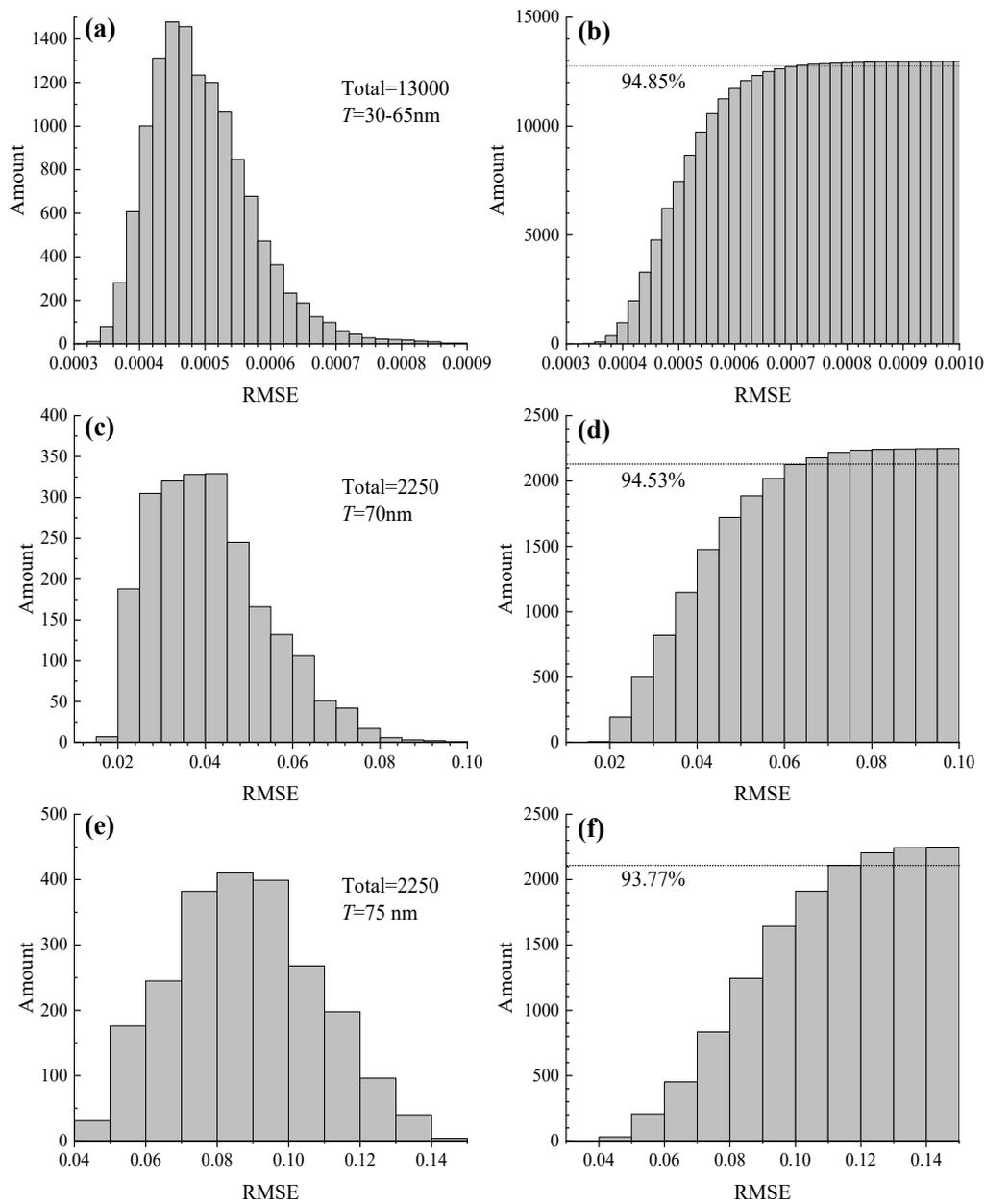

Fig. 10 (a),(c) and (e) The RMSE distribution of the predicted results by the machine learning method; (b),(d) and (f) the cumulative distribution function of the corresponding RMSE in (a),(c) and (e), respectively.

## 4. Conclusion

This paper describes the principle of extension of MBL database by the neural network algorithm. A Monte Carlo simulation method is firstly applied to calculate secondary electron linescan curve for a given set of sample geometry structure parameters and electron beam parameters. By introducing machine learning approach to the Monte

Carlo established MBL database in a small size and taking the Au single-layer and Si double-layer trapezoidal line structures as example, we have verified that that the machine learning approach can predict the MBL database curves in a high accuracy. Hence, the machine learning approach can be employed in practice to extend the MBL database to a much greater size but with the negligible computational cost. This not only saves the huge calculation time but also saves greatly the storage space of MBL database while still keeping the matching accuracy. This work has thus laid the solid foundation for practical application of MBL approach to CD measurement.

**Acknowledgement**

H. Miao acknowledges the National Natural Science Foundation of China (11972235). Y. B. Zou acknowledges the Natural Science Foundation of Xinjinag Uygur Autonomous Region (Grant No. 2022D01A223). S. F. Mao acknowledges the National MCF Energy R&D Program of China (Grant No. 2019YFE03080500) and the Collaborative Innovation Program of Hefei Science Center, CAS (Grant No. 2022HSC-CIP010). Z. J. Ding acknowledges the "111 Project 2.0" Program of Chinese Education Ministry (Grant No. BP0719016). We thank Prof. H. M. Li and the supercomputing centre of USTC for the support of parallel computing.